\documentclass[%
aps,
superscriptaddress,
amsmath,amssymb,
reprint,%
]{revtex4-2}
\usepackage{graphicx}
\usepackage{dcolumn}
\usepackage{bm}
\usepackage[mathlines]{lineno}
\usepackage[utf8]{inputenc}
\usepackage[T1]{fontenc}
\usepackage{mathptmx}
\usepackage[colorlinks,
linkcolor=blue,
anchorcolor=blue,
citecolor=blue,
urlcolor=blue
]{hyperref}
\usepackage{pifont}

\begin{document}

\preprint{AIP/123-QED}

\title{Energy loss enhancement of very intense proton beams in dense matter due to the beam-density effect}

\author{Benzheng Chen\href{https://orcid.org/0000-0002-5669-0800}}
\thanks{These authors contributed equally to this work.}
\affiliation{MOE Key Laboratory for Nonequilibrium Synthesis and Modulation of Condensed Matter, School of Physics, Xi'an Jiaotong University, Xi'an 710049, China}
\author{Jieru Ren}
\thanks{These authors contributed equally to this work.}
\affiliation{MOE Key Laboratory for Nonequilibrium Synthesis and Modulation of Condensed Matter, School of Physics, Xi'an Jiaotong University, Xi'an 710049, China}
\author{Zhigang Deng}
\thanks{These authors contributed equally to this work.}
\affiliation{Science and Technology on Plasma Physics Laboratory, Laser Fusion Research Center, China Academy of Engineering Physics, Mianyang 621900, China}
\author{Wei Qi}
\affiliation{Science and Technology on Plasma Physics Laboratory, Laser Fusion Research Center, China Academy of Engineering Physics, Mianyang 621900, China}
\author{Zhongmin Hu}
\affiliation{MOE Key Laboratory for Nonequilibrium Synthesis and Modulation of Condensed Matter, School of Physics, Xi'an Jiaotong University, Xi'an 710049, China}
\author{Bubo Ma}
\affiliation{MOE Key Laboratory for Nonequilibrium Synthesis and Modulation of Condensed Matter, School of Physics, Xi'an Jiaotong University, Xi'an 710049, China}
\author{Xing Wang}
\affiliation{MOE Key Laboratory for Nonequilibrium Synthesis and Modulation of Condensed Matter, School of Physics, Xi'an Jiaotong University, Xi'an 710049, China}
\author{Shuai Yin}
\affiliation{MOE Key Laboratory for Nonequilibrium Synthesis and Modulation of Condensed Matter, School of Physics, Xi'an Jiaotong University, Xi'an 710049, China}
\author{Jianhua Feng}
\affiliation{MOE Key Laboratory for Nonequilibrium Synthesis and Modulation of Condensed Matter, School of Physics, Xi'an Jiaotong University, Xi'an 710049, China}
\author{Wei Liu}
\affiliation{MOE Key Laboratory for Nonequilibrium Synthesis and Modulation of Condensed Matter, School of Physics, Xi'an Jiaotong University, Xi'an 710049, China}
\affiliation{Xi'an Technological University, Xi'an 710021, China}
\author{Zhongfeng Xu}
\affiliation{MOE Key Laboratory for Nonequilibrium Synthesis and Modulation of Condensed Matter, School of Physics, Xi'an Jiaotong University, Xi'an 710049, China}
\author{Dieter H. H. Hoffmann\href{https://orcid.org/0000-0003-0922-7802}}
\affiliation{MOE Key Laboratory for Nonequilibrium Synthesis and Modulation of Condensed Matter, School of Physics, Xi'an Jiaotong University, Xi'an 710049, China}
\author{Shaoyi Wang}
\affiliation{Science and Technology on Plasma Physics Laboratory, Laser Fusion Research Center, China Academy of Engineering Physics, Mianyang 621900, China}
\author{Quanping Fan}
\affiliation{Science and Technology on Plasma Physics Laboratory, Laser Fusion Research Center, China Academy of Engineering Physics, Mianyang 621900, China}
\author{Bo Cui}
\affiliation{Science and Technology on Plasma Physics Laboratory, Laser Fusion Research Center, China Academy of Engineering Physics, Mianyang 621900, China}
\author{Shukai He}
\affiliation{Science and Technology on Plasma Physics Laboratory, Laser Fusion Research Center, China Academy of Engineering Physics, Mianyang 621900, China}
\author{Zhurong Cao}
\affiliation{Science and Technology on Plasma Physics Laboratory, Laser Fusion Research Center, China Academy of Engineering Physics, Mianyang 621900, China}
\author{Zongqing Zhao}
\affiliation{Science and Technology on Plasma Physics Laboratory, Laser Fusion Research Center, China Academy of Engineering Physics, Mianyang 621900, China}
\author{Leifeng Cao}
\affiliation{Science and Technology on Plasma Physics Laboratory, Laser Fusion Research Center, China Academy of Engineering Physics, Mianyang 621900, China}
\author{Yuqiu Gu\href{https://orcid.org/0000-0001-6979-4556}}
\affiliation{Science and Technology on Plasma Physics Laboratory, Laser Fusion Research Center, China Academy of Engineering Physics, Mianyang 621900, China}
\author{Shaoping Zhu}
\affiliation{Science and Technology on Plasma Physics Laboratory, Laser Fusion Research Center, China Academy of Engineering Physics, Mianyang 621900, China}
\affiliation{Institute of Applied Physics and Computational Mathematics, Beijing 100094, China}
\affiliation{Graduate School, China Academy of Engineering Physics, Beijing 100088, China}
\author{Rui Cheng}
\affiliation{Institute of Modern Physics, Chinese Academy of Sciences, Lanzhou 710049, China}
\author{Xianming Zhou}
\affiliation{MOE Key Laboratory for Nonequilibrium Synthesis and Modulation of Condensed Matter, School of Physics, Xi'an Jiaotong University, Xi'an 710049, China}
\affiliation{Xianyang Normal University, Xianyang 712000, China}
\author{Guoqing Xiao}
\affiliation{Institute of Modern Physics, Chinese Academy of Sciences, Lanzhou 710049, China}
\author{Hongwei Zhao}
\affiliation{Institute of Modern Physics, Chinese Academy of Sciences, Lanzhou 710049, China}
\author{Yihang Zhang}
\affiliation{Beijing National Laboratory for Condensed Matter Physics, Institute of Physics, Chinese Academy of Sciences, Beijing 100190, China}
\affiliation{School of Physical Sciences, University of Chinese Academy of Sciences, Beijing 100049, China}
\author{Zhe Zhang\href{https://orcid.org/0000-0001-8076-5094}}
\affiliation{Beijing National Laboratory for Condensed Matter Physics, Institute of Physics, Chinese Academy of Sciences, Beijing 100190, China}
\affiliation{School of Physical Sciences, University of Chinese Academy of Sciences, Beijing 100049, China}
\author{Yutong Li}
\affiliation{Beijing National Laboratory for Condensed Matter Physics, Institute of Physics, Chinese Academy of Sciences, Beijing 100190, China}
\affiliation{School of Physical Sciences, University of Chinese Academy of Sciences, Beijing 100049, China}
\author{Weimin Zhou\href{https://orcid.org/0000-0001-5474-7511}}
\email{zhouwm@caep.cn}
\affiliation{Science and Technology on Plasma Physics Laboratory, Laser Fusion Research Center, China Academy of Engineering Physics, Mianyang 621900, China}
\author{Yongtao Zhao\href{https://orcid.org/0000-0002-7834-1601}}
\email{zhaoyongtao@xjtu.edu.cn}
\affiliation{MOE Key Laboratory for Nonequilibrium Synthesis and Modulation of Condensed Matter, School of Physics, Xi'an Jiaotong University, Xi'an 710049, China}

\date{\today}

\begin{abstract}
	Thoroughly understanding the transport and energy loss of intense ion beams in dense matter is essential for high-energy-density physics and inertial confinement fusion. Here, we report a stopping power experiment with a high-intensity laser-driven proton beam in cold, dense matter. The measured energy loss is one order of magnitude higher than the expectation of individual particle stopping models. We attribute this finding to the proximity of beam ions to each other, which is usually insignificant for relatively-low-current beams from classical accelerators. The ionization of the cold target by the intense ion beam is important for the stopping power calculation and has been considered using proper ionization cross section data. Final theoretical values agree well with the experimental results. Additionally, we extend the stopping power calculation for intense ion beams to plasma scenario based on Ohm's law. Both the proximity- and the Ohmic effect can enhance the energy loss of intense beams in dense matter, which are also summarized as the beam-density effect. This finding is useful for the stopping power estimation of intense beams and significant to fast ignition fusion driven by intense ion beams.
\end{abstract}
\maketitle

\section{Introduction}
\label{Sec1}

The energy loss of charged particles in matter has been, and continues to be, a key research topic because of its fundamental importance for a large number of applications such as inertial confinement fusion (ICF)~\cite{Li1993_2,Li2006,Hurricane2016}, medicine~\cite{Durante2010,Rackwitz2019}, and astrophysics~\cite{Spitkovsky2008,Huntington2015}. In heavy ion fusion (HIF)~\cite{Bohne1982,Hofmann2018}, $10^{15}$ Uranium ions at about 10~GeV and compressed into a pulse of 15~ns were supposed to be sufficient for ignition. Even at this intensity theories predicted classical beam transport and classical energy deposition where binary collisions dominate the energy deposition process. Laser-generated proton beams today do even exceed this number in the ps regime. Especially fusion scenarios with fast ignition (FI) depend critically on the energy deposition of intense particle beams in matter~\cite{Sarri2010,Guskov2013}. The stopping process of a single charged particle in matter was thoroughly studied by Bethe~\cite{Bethe1930,Bethe1932} and Bloch~\cite{Bloch1933}. Subsequently more effort was devoted to theoretical models~\cite{Li1993,Deutsch2016,Brown2005,Gericke2002} which explained and predicted the energy loss with high precision for a large existing database~\cite{Frenje2019,Zylstra2015,Cayzac2017}. Those experiments were performed over a wide range of energies with low-density ion beams from conventional accelerators. The energy loss of individual particles in matter can be well calculated using these models.

Both experimental and theoretical examination have confirmed that the increase of the ion beam density results in orders of magnitude enhancement of beam stopping, compared to the results by individual particle stopping models. Yet there is few practicable models to calculate this enhanced stopping power.
Deviations of energy loss from individual particle stopping results were first reported by Brandt, Ratkowski and Ritchie~\cite{Brandt1974}, who measured stopping powers for particle clusters of H${_{2}}^{+}$ and H${_{3}}^{+}$ at energies of $\sim$100-keV/u in carbon and gold. This finding was attributed to what they called the vicinage effect of cluster particles and was caused by the wakes of electron-density fluctuations trailing the ions. This prompted a series of investigations leading to improvement and modification of wake theories~\cite{Brandt1976,Marinkovic2015,Nandi2013,Despoja2012} and the vicinage effect~\cite{Basbas1982,Koval2017,LHoir2012,Shubeita2011}. 
There was also study on the transport of actual high-current charged particles in matter, started with intense electron beams~\cite{Logan1974,Wallis1975,Vauzour2012,Guskov2022}.
McCorkle and Iafrate applied the wake theory and vicinage effect of clusters to dense charged particle beams and proposed the beam-density effect for particle stopping~\cite{McCorkle1977}. 
Another important collective effect on the beam energy loss is the Ohmic loss~\cite{Wu2019,Chen2020,Ren2020,Kim2016,Chen2022}, which refers to a decelerating resistive electric field caused by the neutralization of charge and current, in condition of enough free electrons (usually in ionized matter).

Nevertheless, there are still obstacles for the detailed understanding and reliable calculation of energy loss enhancement of intense ion beams in cold targets with few free electrons, in which case the Ohmic effect is weak. The beam-density theory by McCorkle and Iafrate treats the energy loss enhancement from free and bound electrons in the same way, while they should be dealt differently. Additionally, the beam-density theory requires free electrons of non-zero density existing in matter, otherwise it will be invalid due to an infinite result in stopping power calculation. The challenge is to find a proper way to estimate the free electron density of cold matter under intense ion beam impact. A beam-target interaction experiment is also necessary to verify the according results.

In this paper, we report the enhanced stopping power of a cold target for an intense proton beam observed in the experiment. The energy loss of beam protons is measured with high precision and is orders of magnitude higher than the individual particle result, which is attributed to the proximity of beam ions to each other. The local density of free electrons is determined by taking the ionization of the cold target into consideration using proper ionization cross section data. The dependence of the collisional stopping power of dense matter on the beam density is illustrated. Finally, the Ohmic effect is discussed and, together with the proximity effect, incorporated into the beam-density effect.

\section{Experimental results}
\label{Sec2}

\begin{figure}[htb]
	\includegraphics[width=8.5 cm]{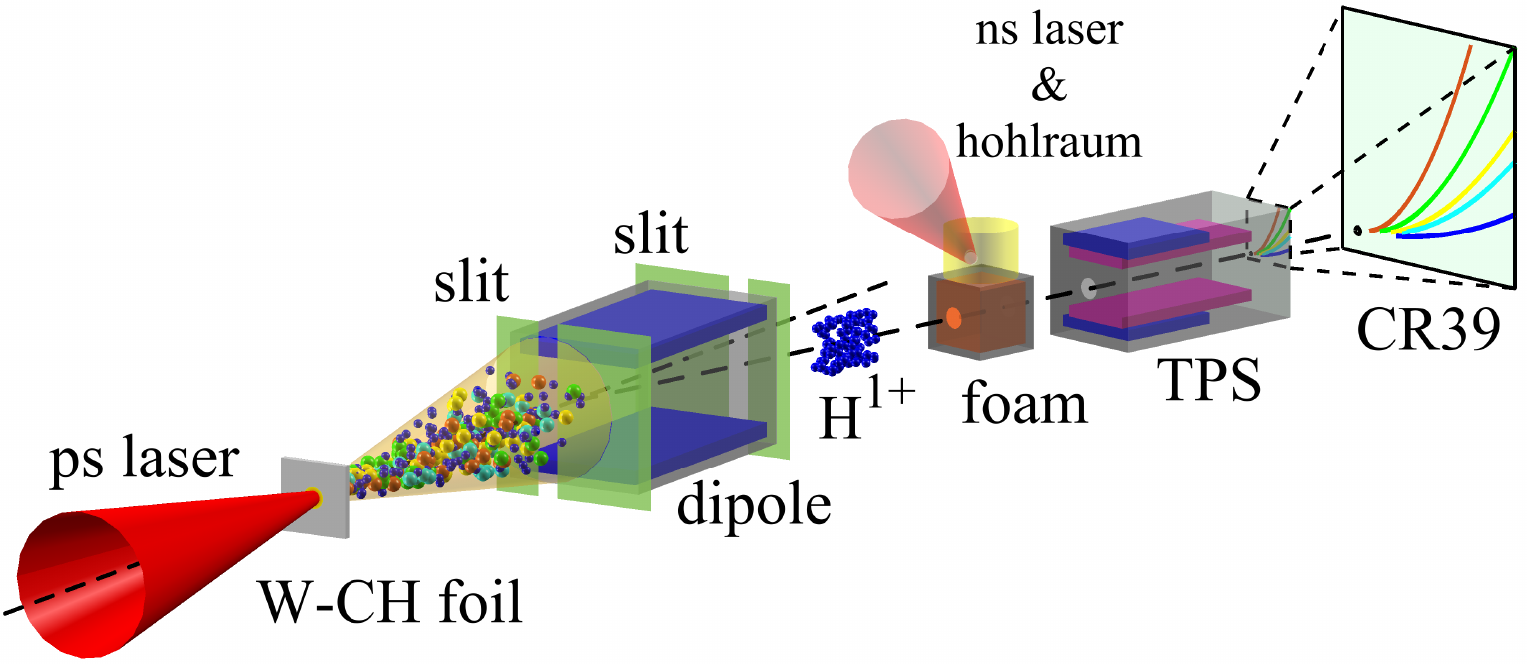}
	\caption{\label{figure1}Schematic set-up of the experimental arrangement including the production of beam particles by TNSA, the selection of quasi-monoenergetic ions, the transport region, and the diagnostic part.
	}
\end{figure}

The experiment of proton beam energy loss in matter was carried out at the XG-III laser facility of Laser Fusion Research Center in Mianyang. Figure~\ref{figure1} shows the schematic set-up of the experimental arrangement. A beam consisting of protons (H$^{1+}$) and carbon ions with different charge states (mostly C$^{1+}$, C$^{2+}$, C$^{3+}$, and C$^{4+}$) is generated by a short and intense laser incident on a CH-coated tungsten foil (15-$\mu$m thick). The laser beam has a duration of 800~fs, focal spot of 20~$\mu$m, and total energy of 150~J. Due to the target normal sheath acceleration (TNSA) mechanism, the mixed beam originates from the rear side of the target with a relative wide spectrum for each kind of the ions. In order to achieve a quasi-monoenergetic beam we used a 500-$\mu$m entrance slit, a magnetic dipole, and another 500-$\mu$m exit slit to select ions according to their momentum ($\boldsymbol{p}$) to charge ($q$) ratio $\boldsymbol{p}/q$.

Protons are the first to arrive at the dipole, since their charge-to-mass ratio is the highest. The C$^{4+}$ ion pulse trails the protons by 30~ns, followed by C$^{3+}$, C$^{2+}$, and C$^{1+}$, which are delayed even more. After the dipole, ions of different species and charge states are separated, and the selected beam pulses are considered quasi-monoenergetic. The production and selection of beam particles ensure that the proton pulse is ahead of carbon pulses and the stopping of protons in the target is unaffected by subsequent carbon ions.

The target of the experiment is a 1-mm-thick C$_9$H$_{16}$O$_8$ foam of 2~mg/cm$^{3}$ density behind a 1-mm aperture.
Placed next to the foam is a gold hohlraum converter, which can generate soft X-rays by irradiating the hohlraum walls with a ns laser pulse (150~J). The soft X-rays then heats the foam and generates homogeneous, ns-long-living, and quasi-static ionized matter. In this way, the influence of electromagnetic fields from the foam heating on beam particles can be ignored. This method of plasma generation has been very well investigated~\cite{Rosmej2011,Rosmej2015,Faik2014}. By turning the ns laser on or off, one can decide whether it is plasma or a cold target.
The following diagnostic part consists of a Thomson parabola spectrometer (TPS)~\cite{Zhang2018,Jung2011,Rajeev2011} and a plastic track detector CR39, which is designed for energy measurement of charged particles. The electromagnetic fields of the TPS separate particles of different charge-to-mass ratios and energies. By analyzing the etched tracks on CR39, we obtain the spectra of particles.

\begin{figure}[htb]
	\includegraphics[width=8.5 cm]{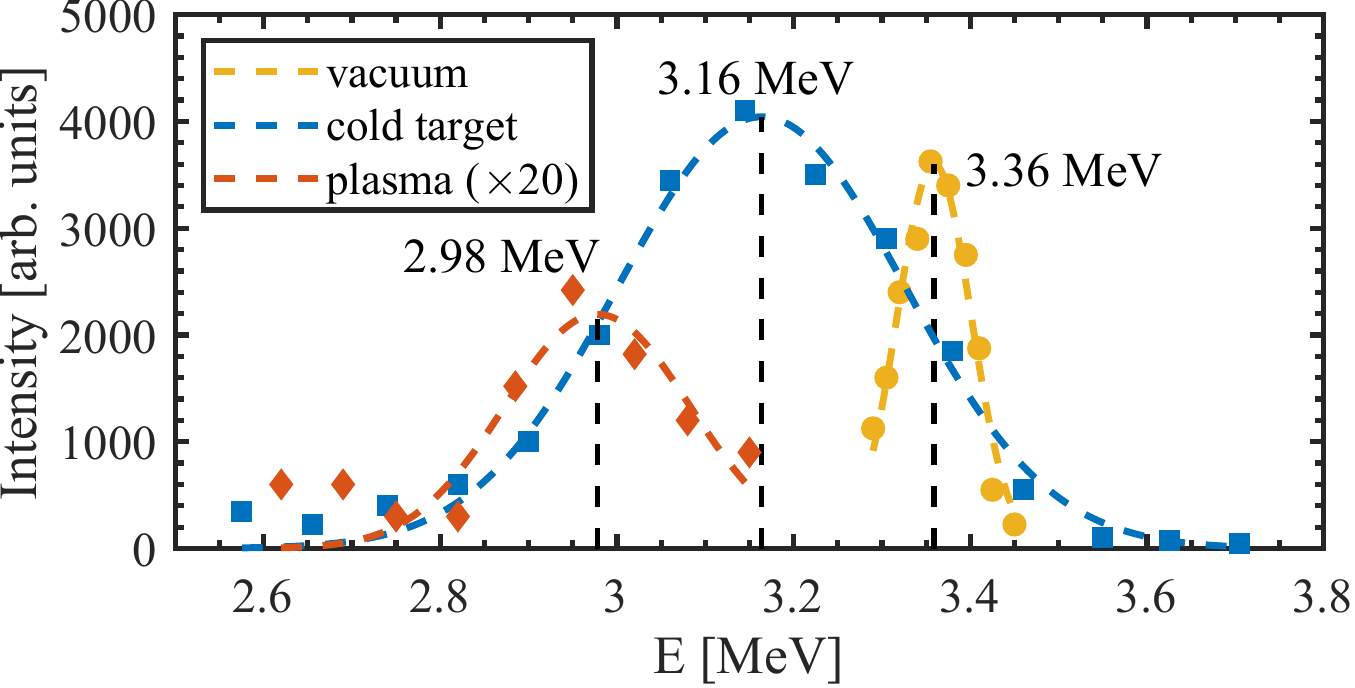}
	\caption{\label{figure2}Energy spectra of the proton particles through vacuum, a cold target, and plasma measured in experiments. The intensity through plasma is magnified 20 times.
	}
\end{figure}

For the proton beam we are interested in, its normalized energy distributions through vacuum and a cold target are shown in Fig.~\ref{figure2}. At an initial injected energy of $E=3.36$~MeV, the TPS resolution is $E/\delta E\sim34$, where $\delta E$ is the energy range covered by the detected beam spot on the CR39. The full width at half maximum (FWHM) of the injected beam is 0.10~MeV. The central energy of the output beam after a cold target is 3.16~MeV, with the FWHM of 0.39~MeV. Experiment shows an energy loss of 0.20~MeV over 1~mm, which is one order of magnitude higher than the theoretical result of 0.021~MeV/mm calculated by Bethe formula~\cite{Bethe1930,Bethe1932}.

In previous work~\cite{Ren2020}, a high degree of proton stopping was observed in plasma as shown in Fig.~\ref{figure2}. The theoretical stopping power of the C$_9$H$_{16}$O$_8$ plasma target with ionization degree of 0.64 ($n_e=4\times10^{20}$~cm$^{-3}$) for 3.36-MeV protons is 0.032~MeV/mm calculated by Bethe formula, with a free electron contribution around 84\%. However, after the beam-plasma interaction, the experimental central energy of the output beam is 2.98~MeV, with a FWHM of 0.25~MeV. This large energy loss of 0.38~MeV/mm was attributed to the strong decelerating electric field caused by the beam-driven return current. PIC simulations indicate that the intensity of this decelerating electric field approaches 1~GV/m. However, the electron density of the cold target is much lower than that of the plasma, and is usually even smaller than the beam density. The corresponding return current is not strong enough to neutralize the beam current or induce the decelerating electric field and therefore can almost be neglected. A different mechanism is required to explain the energy loss enhancement in the cold target.

\section{Beam-density effect on energy loss proposed by McCorkle and Iafrate}
\label{Sec3}

The beam-density effect, considered by McCorkle and Iafrate~\cite{McCorkle1977}, refers to the stopping power enhancement of a dense ion beam in matter, which can reach orders of magnitude when compared to the stopping power of independent particles. This effect is closely related to the collective behavior of the dense beam, where neighboring ions jointly influence the interaction in a coherent way which is called the proximity effect in this paper. To be more specific, the enhanced collisional stopping power for a beam of density $n_b$ in matter is
\begin{align}
	\label{E1}
	S_c&=S_0+S_p\notag\\
	&=S_0\left[1+\int_{0}^{\infty}g\left(r/a\right)4\pi r^2n_b\,\mathrm{d}r\right]\notag\\
	&\approx S_0\left(1+2\pi n_ba^3/3\right)\notag\\
	&=S_0\left(1+N_c\right),
\end{align}
where $S_0$ is the single-particle stopping power, $S_p$ is the stopping power resulting from the proximity effect, and $a\equiv v/\omega_p$. Here, $v$ is the speed of beam particles, $\omega_p=(4\pi n_{fe}e^2/m_e)^{1/2}$ is the electron plasma frequency, $n_{fe}$ is the free electron density, $m_e$ is the electron mass. The interference term $g(r/a)$ reflects the influence of neighboring beam particles and is specified by
\begin{align}
	\label{E2}
	g\left(r/a\right)\approx\begin{cases}
		1/2, &r\leq a;\\
		0, &r>a.
	\end{cases}
\end{align}
The quantity $2\pi n_ba^3/3$ is referred to as the cooperation number $N_c$, which reflects the degree of the proximity effect. The total collisional stopping power is comprised of the single-particle part $S_0$ and the part $S_p$ arising from the proximity effect. For intense ion beams, $S_p$ can be by far larger than $S_0$.

\section{Modification of beam-density effect theory in cold targets}
\label{Sec4}

\subsection{Introducing local ionization degree into beam-density effect theory for cold targets}

It should be noted that the above theory ignores the influence of possible decelerating electric field caused by the beam-driven return current, and underestimates the stopping power in plasma when compared with experimental results (see Appendix~\ref{A1} for details).
Nevertheless, the corresponding formula Eq.~(\ref{E1}) can still be applied to plasma condition in order to estimate its collisional stopping power enhanced by the proximity effect. In the cold target, there are much less free electrons for the neutralization of beam particles. Taking a small value near zero as the free electron density when calculating the values of $a$ and the cooperation number $N_c$  leads to an abnormally high beam-density effect (see Appendix~\ref{A2} for details). To explain the beam-density effect in cold targets and estimate its energy loss enhancement, ionization of the cold targets by intense ion beams must be considered.


Theoretically, protons with energy larger than the binding energy of bound electrons can excite bound target electrons to higher orbits and even ionize them, especially the valence electrons with low binding energy. In cold targets, nearly all the free electrons arise from the ionization by protons and these ionized electrons must be considered. The density of local ionized free electrons is $n_{lfe}=\alpha_ln_{te}$, where $\alpha_l$ is the local ionization degree, $n_{te}$ is the total electron density. These electrons lead to $\omega_p=(4\pi n_{lfe}e^2/m_e)^{1/2}$ and can contribute to the interference term in Eq.~(\ref{E2}) by influencing the value of $a=v/\omega_p$. For the C$_9$H$_{16}$O$_8$ target in the experiment, the total electron density is $n_{te}=6.4\times10^{20}$~cm$^{-3}$. Additionally, the bound electrons represent a large fraction of the total electrons. Their resonance frequency is related to the ionization energy $\omega_k=I_k/\hbar$. By summing up the contributions of free electrons $S_{fe}$ and bound electrons $S_{be}$, the total collisional stopping power for a dense beam in matter is rewritten as
\begin{align}
	\label{E3}
	S_{c}=&S_{fe}+S_{be}\notag\\
	=&S_{f0}\left[1+\frac{2}{3}\pi n_b\left(\frac{v}{\omega_p}\right)^3\right]+\sum_k S_{b0k}\left[1+\frac{2}{3}\pi n_b\left(\frac{v}{\omega_k}\right)^3\right]\notag\\
	=&\left(S_{f0}+\sum_k S_{b0k}\right)+\left[S_{f0}\frac{n_b}{12\sqrt{\pi}}\left(\frac{m_ev^2}{\alpha_ln_{te}e^2}\right)^{3/2}\right.\notag\\
	&\left.+\sum_k S_{b0k}\cdot\frac{2}{3}\pi n_b\left(\frac{v}{I_k/\hbar}\right)^3\right]\notag\\
	=&S_{0}+S_p.
\end{align}
Here $S_{f0}$ and $S_{b0k}$ are the stopping power for an individual proton resulting from free electrons and bound electrons, respectively. According to Bethe formula~\cite{Bethe1930,Bethe1932} with thermal correction~\cite{Chandrasekhar1943}, they are the components of $S_0$ and can be specified by
\begin{align}
	\label{E4}
	S_{f0}=\frac{4\pi e^4Z^2}{m_ev^2}G\left(\frac{v}{v_{th}}\right)n_{lfe}{\rm ln}\left(\frac{2m_ev^2}{\hbar\omega_p}\right),
\end{align}
\begin{align}
	\label{E5}
	S_{b0k}=\frac{4\pi e^4Z^2}{m_ev^2}n_{bek}{\rm ln}\left(\frac{2m_ev^2}{I_k}\right).
\end{align}
Here, $Z$ is the effective charge state of injected ion beams, $n_{lfe}$ and $n_{bek}$ are the densities of free and bound electrons. $G(x)=\mathrm{erf}(\sqrt{x})-2\sqrt{x/\pi}\exp(-x)$ is the Chandrasekhar thermal correction factor, which approaches 1 when $x\gg$1.

\subsection{Determining local ionization degree through cross sections}

The ionization of bound electrons in cold targets does also depend on the beam density. In order to determine their contribution to $S_{c}(n_b)$, the cross section for the target ionization by protons needs to be investigated.

By incorporating several adjustments into the plane-wave Born approximation (PWBA)~\cite{Merzbacher1958,Benka1978} theory, including approximation of the perturbed stationary state (PSS) and corrections for energy loss (E), Coulomb deflection (C), and relativistic (R) effects, Brandt and Lapicki~\cite{Brandt1979,Brandt1981} proposed the ECPSSR theory, which has proven to be one of the most successful theories for evaluating cross sections of $K$- and $L$-shell ionization for ion impact on targets. Here, a versatile and fast program developed by Liu and Cipolla, ISICS~\cite{Liu1996,Batic2013}, is used to calculate ionization cross sections based on ECPSSR theory. The output cross sections for $K$- and $L$-shell ionization in H, C, O atoms by 3.36-MeV protons are shown in Table~\ref{table1}.

\begin{table}[htb]
	\renewcommand\arraystretch{1.3}
	\caption{\label{table1}%
		 $K$- and $L$-shell ionization cross sections of H, C, O atoms in barns (1 barn = $10^{-24}$~cm$^{2}$) for 3.36-MeV protons calculated by ISICS.}
	\begin{ruledtabular}
		\begin{tabular}{ccccc}
			Elements & $\sigma_K$ & $\sigma_{L_1}$ & $\sigma_{L_2}$ & $\sigma_{L_3}$\\
			\hline
			H &1.384$\times$10$^{7}$ & & &\\
			C &4.904$\times$10$^{5}$ & &1.384$\times$10$^{7}$ &8.484$\times$10$^{7}$\\
			O &1.929$\times$10$^{5}$ &8.778$\times$10$^{6}$ &1.968$\times$10$^{8}$ &3.937$\times$10$^{8}$\\
		\end{tabular}
	\end{ruledtabular}
\end{table}

The number of electrons ionized by the impact of a proton on a cold target of length $d$ is specified by
\begin{align}
	\label{E6}
	N_{ie}=\left(\sum_k \sigma_k n_k\right)d,
\end{align}
where $n_k$ is the number density of electrons with cross section of $\sigma_k$. For a proton beam with radium $r=0.5$~mm and beam length $L=25~\mu$m, the number of protons contained in this range is $n_b\pi r^2L$. The local ionization degree $\alpha_l$ is related to cross sections by
\begin{align}
	\label{E7}
	\alpha_l=\frac{N_{ie}\left(n_b\pi r^2L\right)}{n_{te}\pi r^2d}=\left(\sum_k \sigma_k n_k\right)\frac{n_b}{n_{te}}L.
\end{align}
Substituting the data of Table~\ref{table1} into Eq.~(\ref{E7}) yields the local ionization degree of the C$_9$H$_{16}$O$_8$ foam
\begin{align}
	\label{E8}
	\alpha_l\approx6.35\times10^{-19}n_b\left({\rm cm}^{-3}\right).
\end{align}

The local ionization degree defined by Eq.~(\ref{E8}) is the minimum value when ionization cross sections of all atoms are considered. In this way, the stopping power divergence (explosion) at low ionization degrees is eliminated.

\section{Calculating the total stopping power}

\begin{figure}[htb]
	\includegraphics[width=8.5 cm]{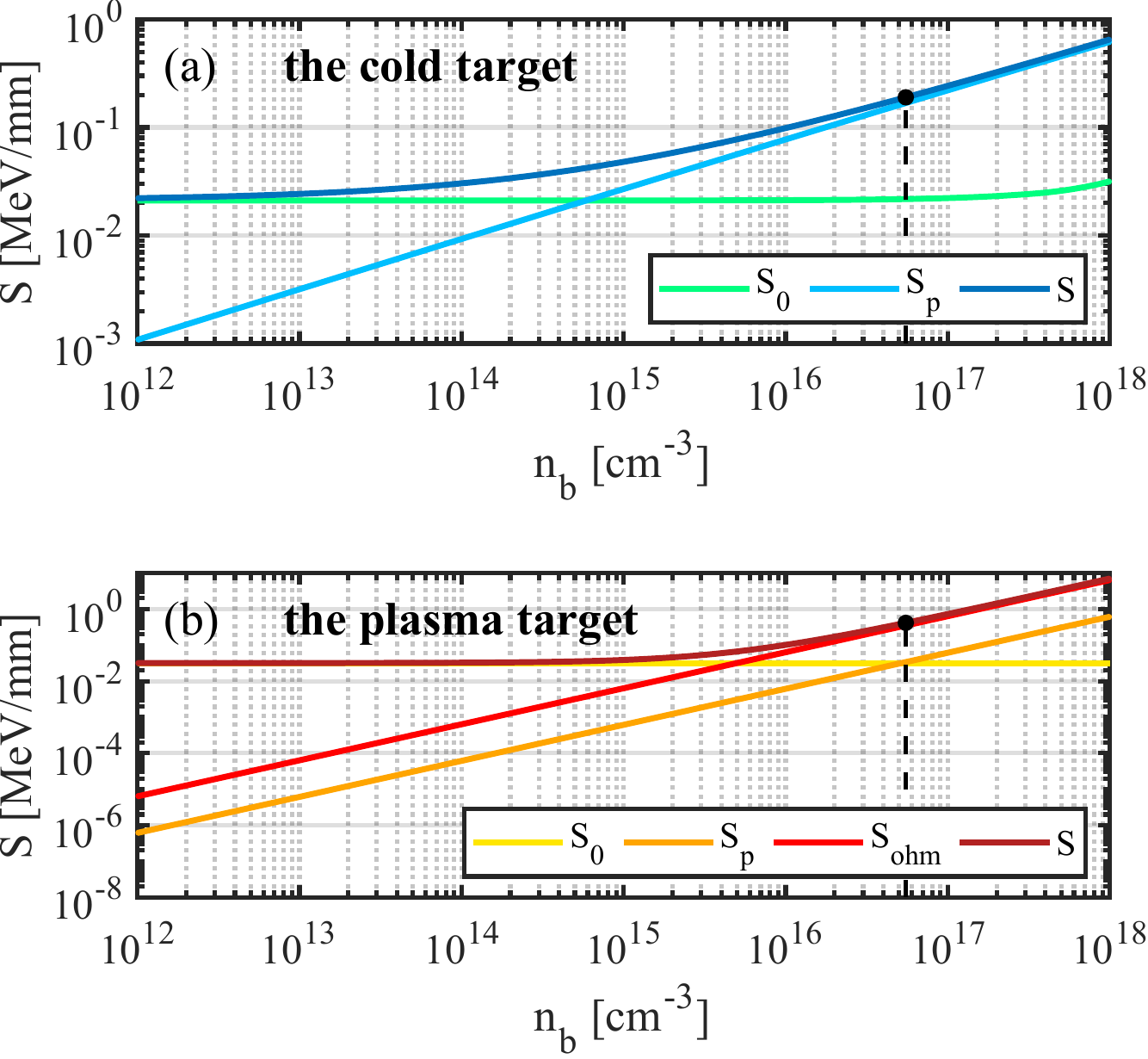}
	\caption{\label{figure3}Calculated stopping power as functions of the proton beam density for (a) a cold target and (b) plasma. At a high beam density, $S_p$ is prominent for the cold target, while $S_{ohm}$ makes up most of the stopping power for plasma.
	}
\end{figure}

Substituting Eqs.~(\ref{E4}), (\ref{E5}) and (\ref{E8}) into Eq.~(\ref{E3}) yields $S_{0}$, $S_{p}$, and the total collisional stopping power $S$ of the cold target as functions of only the beam density $n_{b}$, as displayed in Fig.~\ref{figure3}(a). When the beam density rises from $10^{12}$ to $10^{18}$~cm$^{-3}$, the local ionization degree increases linearly from $6.35\times10^{-7}$ to 0.635. At a low beam density, the beam-density effect is weak and the stopping power is mostly given by the single-particle result. When the beam density rises, the stopping power caused by the proximity effect increases sharply. It becomes comparable to the individual particle stopping power around the beam density of $6\times10^{14}$~cm$^{-3}$, and accounts for nearly 90$\%$ of $S=0.19$~MeV/mm at $n_b=5.5\times10^{16}$~cm$^{-3}$, which is of the same order of density as the simulation estimation~\cite{Ren2020}.

Considering a plasma target, what must be considered is the Ohmic stopping power $S_{ohm} = ZeE$, where $E=\eta J$, $\eta$ is the resistivity and $J$ is the return current according to Ohm's law. The resistivity here is higher than the classical value of $\eta_c=m_e\nu_{ei}/n_{fe}e^2$, where the classical electron-ion collision frequency is specified by
\begin{align}
	\label{E9}
	\nu_{ei} = \frac{\pi Z_i^2n_ie^4{\rm ln}\Lambda}{m_e^{1/2}T_e^{3/2}},
\end{align}
where $Z_i$ is the ionization charge state of background plasmas, ln$\Lambda$ is the coulomb logarithm and $T_e$ is the plasma temperature.
This is because the collision between electrons and injected ions also needs to be considered for the resistivity calculation. According to the theory of wave-particle nonlinear interactions by Sagdeev and Galeev~\cite{Sagdeev1969}, the collision frequency between electrons and the beam is
\begin{align}
	\label{E10}
	\nu_{eb} = \omega_p\frac{v}{\sqrt{T_e/m_e}}\frac{T_b}{T_e},
\end{align}
where the ion beam temperature (average thermal energy) $T_b\approx$18~eV in our case (see Appendix~\ref{A3} for details). The final resistivity is then written as $\eta=m_e(\nu_{ei}+\nu_{eb})/n_{fe}e^2$.

As shown in Fig.~\ref{figure3}(b), the total stopping power in plasma at high beam densities is mainly composed of the Ohmic stopping power $S_{ohm}$, with a resistivity $\eta$ of 0.16~$\Omega\cdot$cm. At a beam density of $5.5\times10^{16}$~cm$^{-3}$, it reaches 0.42~MeV/mm, which is close to the experimental result of 0.38~MeV/mm.

The beam-density effect is an overall collective phenomenon caused by the high number of injected beam particles. It illustrates that the energy loss of intense ion beams in dense matter is composed of individual particle energy loss $S_0$, enhanced collisional energy loss $S_p$, which results from proximity effect of beam particles, and Ohmic energy loss $S_{ohm}$, written as $S=S_0+S_p+S_{ohm}$.
Usually only one of them may be dominant in specific scenarios. If there are few free electrons for neutralization of the beam, the beam-density effect may take the form of proximity effect, and the enhanced collisional energy loss is prominent~\cite{McCorkle1977}. If the return current is strong in plasmas with numerous free electrons, one may put the influence of decelerating electric field (Ohmic field) first~\cite{Ren2020}.

\begin{figure}[htb]
	\includegraphics[width=8.5 cm]{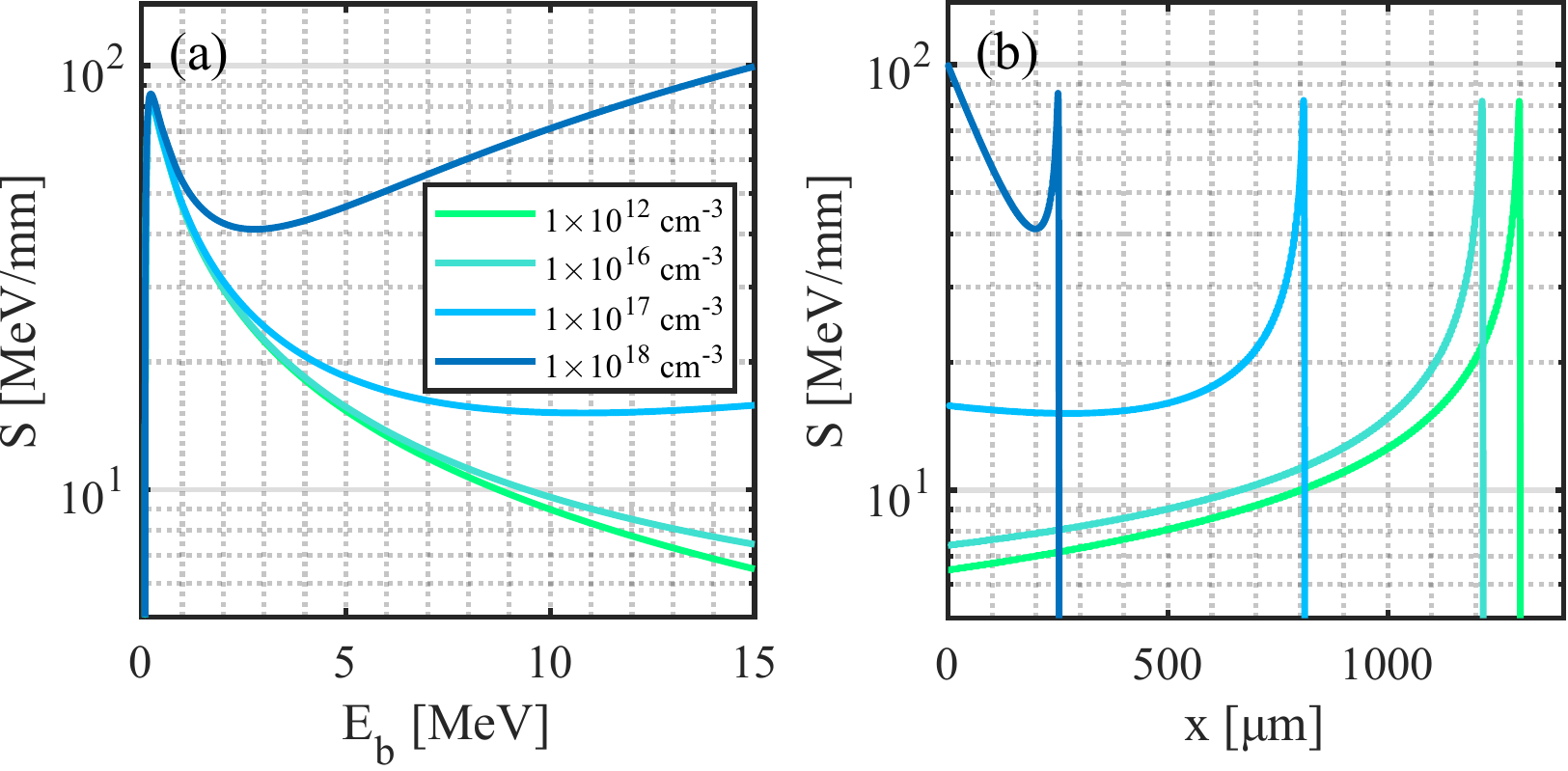}
	\caption{\label{figure4}(a) The calculated proton stopping power in solid aluminum as a function of proton energy with different beam densities, and (b) the corresponding Bragg curves of 15-MeV protons.
	}
\end{figure}

The beam-density effect proposes new requirements for ion-driven warm dense matter (WDM) and fast ignition. The traditional relation between the stopping power and the beam energy can become unreliable for intense ion beams. Figure~\ref{figure4}(a) plots the calculated proton stopping power in solid aluminum as a function of proton energy for beam densities of $10^{12}$, $10^{16}$, $10^{17}$, and $10^{18}$~cm$^{-3}$. There are so many free electrons in solid aluminum that the Ohmic stopping power $S_{ohm}$ is responsible for the stopping power enhancement in this case. Correspondingly, the Bragg peaks of proton energy deposition with initial energy of 15~MeV are shown in Fig.~\ref{figure4}(b). The beam-density effect can greatly affect the particle deposition in dense matter.

\section{Conclusions}
\label{Sec5}

To sum up, a stopping power enhancement of one order of magnitude for an intense proton beam in a cold target was measured. The observed effect is attributed to the proximity effect. Here, a new method for numerically calculating the collisional stopping power for intense ion beams in dense matter is introduced. It modifies the beam-density effect theory with the local ionization caused by the intense beam. By obtaining the ionization cross sections of the cold target and considering Ohmic energy loss, the total stopping power is determined as a function of the beam density. To reach the energy loss of 0.20~MeV over 1~mm in the experiment, our method gives a beam density of $5.5\times10^{16}$~cm$^{-3}$, which agrees with the simulation estimation. The energy loss of intense ion beams in dense matter is composed of individual particle energy loss, enhanced collisional energy loss, which results from proximity effect of beam particles, and Ohmic energy loss. This research provides a new perspective to theoretically calculate the dramatic stopping power enhancement of intense ion beams in dense matter, and proposes new requirements for ICF and high-energy-density physics.

\section*{Data availability}

The data that support the findings of this study are available from the corresponding author upon reasonable request.

\begin{acknowledgments}
	This work was supported by National Key R\&D Program of China (Grant No. 2019YFA0404900), Science Challenge Project (Grant No. TZ2016005), National Natural Science Foundation of China (Grants No. U2030104, No. 11705141, No. 11775282, No. 11605269, and No. U1532263), China Postdoctoral Science Foundation (Grants No. 2017M623145 and No. 2018M643613), and Science and Technology on Plasma Physics Laboratory (Grant No. J202108010).
\end{acknowledgments}

\appendix
\section{Collisional stopping power for a proton beam in plasma by the beam-density theory}
\label{A1}
\begin{figure}[htb]
	\includegraphics[width=8.5 cm]{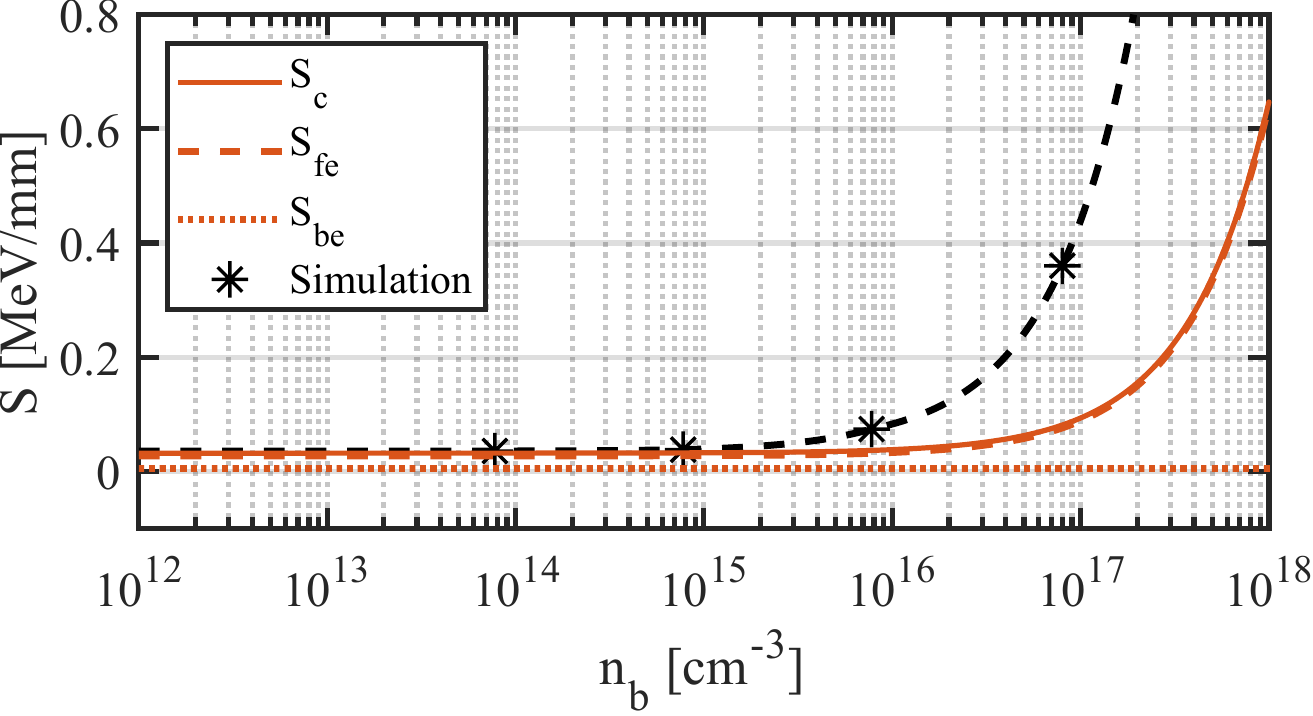}
	\caption{\label{figure5}Several stopping powers for proton beams with densities $10^{12}$--$10^{18}$~cm$^{-3}$ in the plasma of the experiment. The initial beam energy is 3.36~MeV. The curves of $S_{c}$, $S_{fe}$ and $S_{be}$ are outputs of Eq.~(\ref{E3}). Several simulation results are also presented.
	}
\end{figure}

Fig.~\ref{figure5} shows the collisional stopping power $S_{c}$ for a 3.36-MeV proton beam in the plasma of the experiment as a function of the beam density by Eq.~(\ref{E3}), and compares the results with simulations.
The modified collisional stopping power $S_{c}$ is composed of the free electron contribution $S_{fe}$ and the bound electron contribution $S_{be}$. The latter only takes a small percentage and can be regarded as a constant ($\sim$4.9~keV/mm). It is the increase of $S_{fe}$ that makes $S_{c}$ rise sharply.
When the beam density rises from $10^{12}$ to $10^{18}$~cm$^{-3}$, the collisional stopping power in the plasma can increase by one order of magnitude from 0.032 to over 0.6~MeV/mm. Simulations also verify the enhancement of beam energy loss~\cite{Ren2020}. In addition, to reach a stopping power of 0.38~MeV/mm as measured in the plasma of the experiment, simulations and the theory give beam densities of $\sim8\times10^{16}$ and $5\times10^{17}$~cm$^{-3}$, respectively. The beam-density theory apparently underestimates the enhancement of the stopping power in plasma. This is because it calculates the collisional energy loss enhancement resulting from proximity of beam ions, but ignores the impact of the decelerating electric field caused by the beam-driven return current, which is the main reason for the enhanced energy loss in plasma case.

\section{Problems of explosive stopping power increase at low electron densities}
\label{A2}
Applying the beam-density effect theory of McCorkle and Iafrate~\cite{McCorkle1977} requires free electrons of certain density existing in matter. Otherwise the plasma frequency $\omega_{p}$ would be very small for an electron density $n_{fe}$ near zero, leading to enormous values of $a=v/\omega_{p}$ and cooperation number $N_c$ and then the explosion of the stopping power $S_c$. 

\begin{figure}[htb]
	\includegraphics[width=8.5 cm]{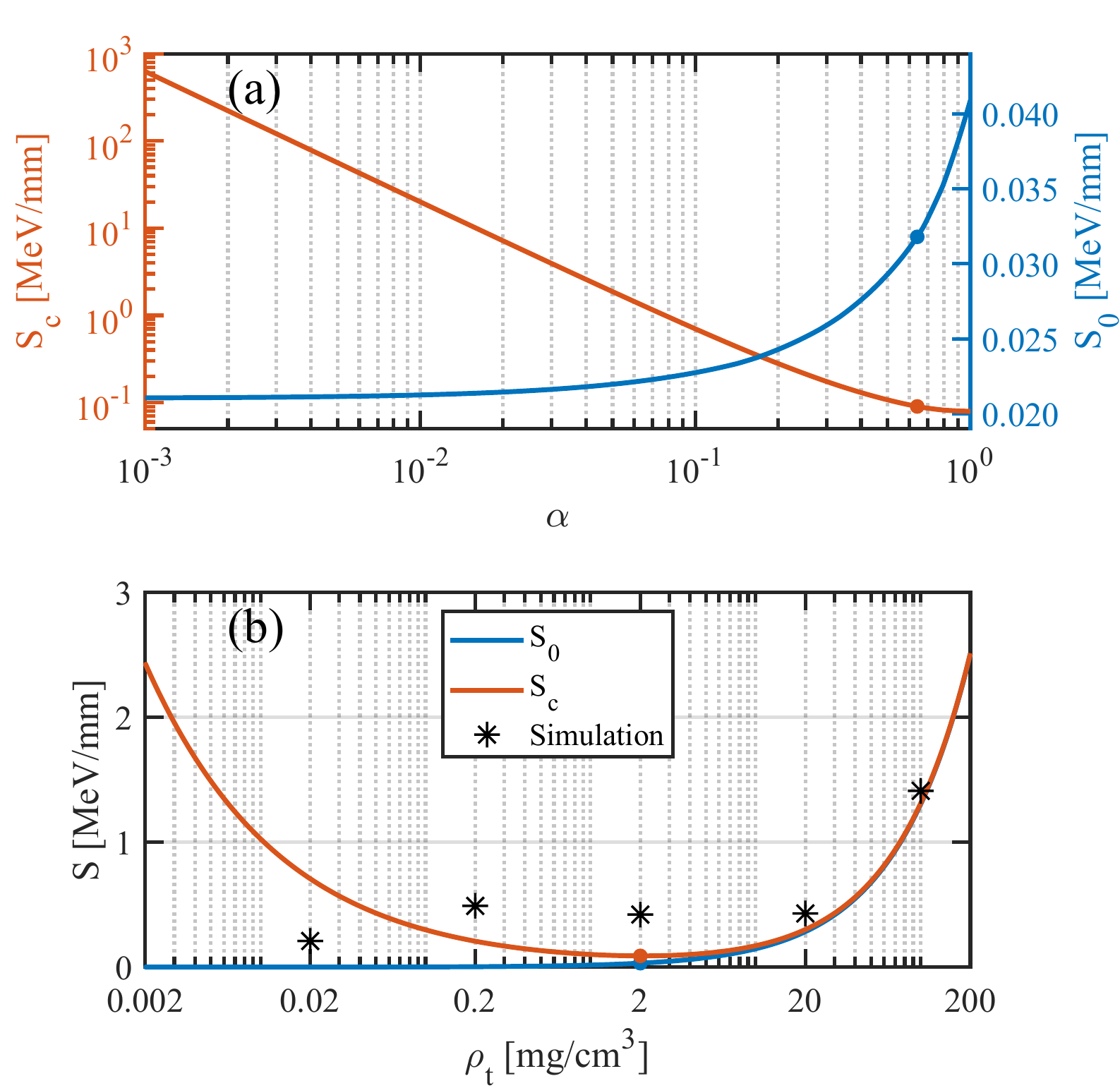}
	\caption{\label{figure6}Explosive stopping power increase of targets with different ionization degrees and different mass densities. The blue and red curves are results for a single proton and for a dense beam with density $n_b=8\times10^{16}$~cm$^{-3}$, respectively. (a) The target has mass density 2~mg/cm$^{3}$ and ionization degree varying from $10^{-3}$ to 1 (fully ionized). (b) The target has the same ionization degree with the experiment target of 0.64. The mass density of the target varies from 0.002 to 200~mg/cm$^{3}$. The initial energy of both the individual particle and the dense beam is 3.36~MeV. Marked dots are the results corresponding to the experiment.}
\end{figure}

Fig.~\ref{figure6} shows this unphysical energy loss increase of dense proton beam in cold targets and low-density plasma by the beam-density effect theory. In Fig.~\ref{figure6}(a), the stopping power of the C$_9$H$_{16}$O$_8$ target with mass density $\rho_t=2~{\rm mg/cm}^{3}$ and ionization degree $\alpha$ varying from $10^{-3}$ to 1 (fully ionized) for a single proton $S_0$ (blue) and a proton beam $S_c$ (red) with density $n_b=8\times10^{16}$~cm$^{-3}$ is presented. The initial beam energy is $E=3.36$~MeV. The energy loss result of the proton beam is based on the original beam-density effect theory and is obtained by Eq.~(\ref{E1}). The local ionization is not considered here. Note that two curves are shown with different $y$ axes and the stopping power with beam-density effect is orders of magnitude higher than that for a single particle. Marked dots in Figs.~\ref{figure6} show the results corresponding to the experimental plasma with temperature 17~eV and ionization degree 0.64.

When the target evolves from plasma to a cold target with the ionization degree $\alpha$ declining from 0.64 to 0.001, though the single-particle stopping power $S_0$ decreases from 0.032 to 0.021~MeV/mm, the total collisional stopping power increases dramatically from 0.090 to $6\times10^2$~MeV/mm according to Eq.~(\ref{E1}). This enhancement of stopping power is far more larger than the experimental result of 0.20~MeV/mm in cold targets. Because 1) an ionization degree of $10^{-3}$ is lower than the actual value and 2) both free and bound electron contributions $S_0$ multiply the same cooperation number $2\pi n_b(v/\omega_p)^3/3$ while the bound electron part should be exclude and time a smaller cooperation number $2\pi n_b(v/\omega_k)^3/3$. By calculating the minimum local ionization degree using ionization cross sections in Eq.~(\ref{E8}), the corresponding collisional stopping power is computed as two parts in Eq.~(\ref{E3}). In this way, the stopping power explosion is eliminated.

Fig.~\ref{figure6}(b) displays $S_0$ and $S_c$ of the C$_9$H$_{16}$O$_8$ target with fixed ionization degree 0.64 (same as the plasma target in the experiment) and mass density $\rho_t$ from 0.002 to 200~mg/cm$^{3}$. Reducing the target density can also make the electron density of the target decline. If Eq.~(\ref{E1}) of the beam-density effect theory is applied to low-density plasma, the stopping power would be abnormally enormous and obviously unphysical. In fact, when the electron density is lower than the beam density, the polarization wakes produced by the beam particles is weak, and so is the collective effect of the beam particles. The lower the target density is, the closer the stopping power to the single-particle case. Simulation results also show the tendency of falling as the target density decreases in the range of $\rho_t<0.2~{\rm mg/cm}^{3}$. It is under the condition that the total electron density $n_{te}$ of the target is higher than the beam density $n_b$ that we consider using the beam-density effect theory to calculate the enhanced collisional stopping power.

\section{Ion beam temperature estimation}
\label{A3}
The ion beam temperature $T_b$ discussed here refers to the average thermal energy of injected protons, which excludes the initial kinetic energy of $E=3.36$~MeV. It mainly results from beam divergence and is restricted to the width of the slit $d'$ behind the dipole and the aperture $d$ in front of the target (see Fig.~\ref{figure1}). Polar coordinates ($r'$, $\theta'$) and ($r$, $\theta$) are used in planes of the slit (plane I) and the target aperture (plane II), respectively. Only protons with $r'$ smaller than $d'/2$ are considered to make the system axisymmetric for simplicity. The area density functions of protons in plane I and plane II are $g(r')$ and $f(r)$. The distance between plane I and plane II is $L$. The average thermal energy is specified by
\begin{align}
	\label{EA1}
	T_b=\frac{\int_{0}^{2\pi}\int_{0}^{\frac{d}{2}}\frac{1}{2}m_p\overline{v_{t}^2}f(r)r\,\mathrm{d}r\mathrm{d}\theta}{\int_{0}^{2\pi}\int_{0}^{\frac{d}{2}}f(r)r\,\mathrm{d}r\mathrm{d}\theta},
\end{align}
where $m_p$ is the proton mass,
\begin{align}
	\label{EA2}
	\overline{v_{t}^2}(r,\theta)=\frac{\int_{0}^{2\pi}\int_{0}^{\frac{d'}{2}}v_t^2g(r')r'\,\mathrm{d}r'\mathrm{d}\theta'}{\int_{0}^{2\pi}\int_{0}^{\frac{d'}{2}}g(r')r'\,\mathrm{d}r'\mathrm{d}\theta'},
\end{align}
$v_t=vl/L$ is the transverse velocity, $v=\sqrt{2E/m_p}$ and $l=\sqrt{r^2+r'^2-2rr'\cos{(\theta-\theta')}}$ is the transverse distance between ($r'$, $\theta'$) and ($r$, $\theta$).

In the experiment, $d'=0.5$~mm, $d=1$~mm, $L=17$~cm. Considering $f(r)$ and $g(r')$ as two constant functions, one obtains $T_b\approx$18~eV. If Gaussian distribution is applied, $f(r)=A\exp{(-r^2/2\sigma^2)}$ and $g(r')=A'\exp{(-r'^2/2\sigma'^2)}$, where $\sigma=d/2\sqrt{2\ln{2}}$  and $\sigma'=d'/2\sqrt{2\ln{2}}$. The according beam temperature is about 16~eV. Since the plasma temperature $T_e$ is 17~eV, it can be assumed that $T_b\approx T_e$ in our experimental condition.


\end{document}